\documentstyle[prl,aps,psfig,floats,twocolumn]{revtex}
\begin{document}
\draft
\twocolumn[\hsize\textwidth\columnwidth\hsize\csname
@twocolumnfalse\endcsname
\title{Interacting Individuals Leading to Zipf's Law}
\author{Matteo Marsili and Yi-Cheng Zhang}
\address{Institut de Physique Th\'eorique, 
Universit\'e de Fribourg, CH-1700}
\date{\today}
\maketitle
\begin{abstract}
We present a general approach to explain the Zipf's law of city
distribution. If the simplest interaction (pairwise) is assumed,
individuals tend to form cities in agreement with the well-known
statistics 
\end{abstract}

]
\narrowtext

\begin{figure}
\centerline{\psfig{file=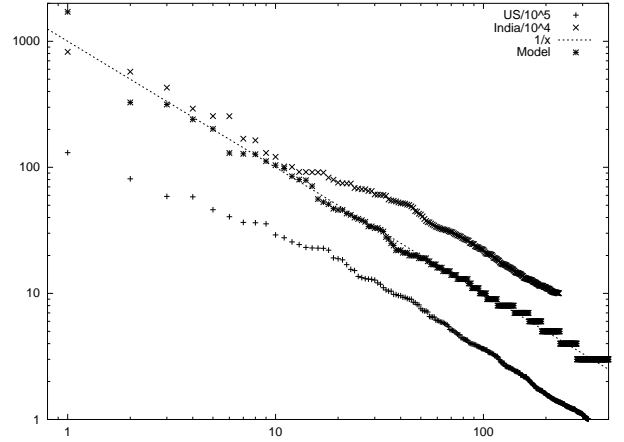,width=8.5cm,angle=270}}
\caption{Zipf plot for cities of population larger
than $10^5$ in the US and in India, and for 
a simulation of the master equation 
with $p=10^{-3}$ and $W=10^6$, $\alpha=2$.}
\label{fig1}
\end{figure}

Zipf\cite{zipf},
half a century ago,  has found that the city sizes obey an 
astonishingly 
simple distribution law, which is attributed to the more 
generic
{\it least effort principle of human behavior}. Let us 
denote by
$q_m$ the number of cities having the population size 
$m$, then
$R(m)=\int_m^\infty q_{m'}~dm'$
defines the rank, i.e. the largest city
has $R=1$, the second largest $R=2$, etc. Zipf found 
that empirically,
$R(m)\sim 1/m^\gamma$, with $\gamma\approx 1$ (see figure
\ref{fig1}). Remarkably Zipf showed that the 
scaling exponent $\gamma=1$ is very close to reality for 
many different societies and during various time periods. 
More recent data \cite{zm,data}
shows some variations from the pure $\gamma =1$ 
result as also shown in figure \ref{fig1}. 
Countries which have a peculiar social structure, such as the former 
USSR or China, do not follow Zipf law. For other 
developed countries Zipf law remains a rather good 
approximation. Such a generic law calls for 
generic explanation, since different countries (e.g. 
Germany and USA) 
have different cultural, economic structures, and their 
people have 
innumerable reasons to choose whether to live in a big or 
small city.
Yet collectively the society self-organises, without 
express
wishes of authorities, to obey the Zipf's law.

Human settlements on the earth's surface appear to be 
clustered, hence cities. This is because individuals {\it 
interact}
with each other through social, economic and cultural 
ties. 
In very primitive times 
an individual (or a family) should perform all basic 
activities to 
survive; there was no need to form a large cluster beyond 
the size of a tribe.
In modern times, ever refined mutual 
cooperation/competition brings people
to live together. Yet this tendency does not seem to 
lead to a single 
mega-city in the world: many of us prefer to live in a 
big city, equally many
may hate and escape it for all its negative impacts.
Somehow the ensuing compromise results in a robust
statistical distribution, the Zipf's law.

All these very general features call for a equally 
general approach to model
city distribution. Below we 
propose
a general framework
using master equations. 
Let there be $Q$ cities and $m_i$ citizens in the 
$i^{\rm th}$ city. The model is defined in terms of a
master equation, assigning transition rates for the
growth $w_a(m_i)$ or decrease $w_d(m_i)$ of the 
population
of a city of size $m_i$. In other words we assume that
with a probability $w_a(m_i)dt$ a new citizen arrives
in city $i$ in the time interval $(t,t+dt)$, so 
that
$m_i\to m_i+1$. With a probability $w_d(m_i)dt$ one
of the $m_i$ citizens departs so that $m_i\to m_i-1$. 
We also assume that there is a small probability $p\,dt$
that a new city is created, with a single citizen.
Assigning the transition rates $w_a(m)$ and $w_d(m)$
specifies the model. 
Note that birth and death are not explicitly
considered, these can be included in the transition 
probabilities. 
Note also that once an individual
leaves a city, not necessarily he/she will settle in 
another city
right away, thus the total number of city dwellers is 
not conserved. 
Thus the whole system is composed of the city dwellers and a 
reservoir of unsettled travellers whose number is unregistered.

One can study this problem introducing the 
average number $q_{m,t}$
of cities of size $m$ at time $t$, which satisfies the 
master equation:
\begin{eqnarray}
\partial_t
q_{m,t}&=&w_d(m+1)q_{m+1,t}-w_d(m)q_{m,t}+p\delta_{m,1}
\nonumber\\
&~&+w_a(m-1)q_{m-1,t}-w_a(m)q_{m,t}.
\end{eqnarray}
The parameters of the equation are the transition rates 
$w_d(m),~p,~w_a(m)$. 

The total number $n(t)=\sum_m m q_{m,t}$ 
of persons and the total number $Q$ of cities
are generally not constant:
\begin{eqnarray}
\partial_t Q&=&p-w_d(1)q_{1,t}\\
\partial_t n&=&p-\sum_{m=1}^\infty [w_d(m)-w_a(m)]q_{m,t}.
\end{eqnarray}

Let us focus on the stationary state solutions $\partial_t 
q_{m,t}=0$, 
for which $q_{m,t}$ is simply denoted $q_{m}$, 
independent of time, and $n$ and $Q$ are  
constant on average.
This leads us to the equation 
\begin{eqnarray}
w_d(m+1)q_{m+1}-w_d(m)q_{m}+p\delta_{m,1}&~&\nonumber\\
+w_a(m-1)q_{m-1}-w_a(m)q_{m}&=&0 ,
\end{eqnarray}
where the rate $w$ are also independent of time.
The problem can be readily handled  with the aid of the 
generating function $g(s)=\sum_m s^m q_m$.

Let us first consider the linear case $w_d(m)=Dm$ and 
$w_a(m)=Am$. This corresponds to independent decisions
by the individuals.
It is convenient to choose the constants to be 
$A=(1-p)/n, D=1/n$, where $n$ counts the average total number 
of citizens (constant). 
In general there is a deficit for each existing 
city $w_d>w_a$, since a departing 
individual have finite chance to create a new city. 
The above equation is readily solved with the result:
\begin{equation}
q_m=\frac{np}{1-p}\frac{(1-p)^m}{m}\sim \frac{1}{m} 
e^{-pm}.
\end{equation}
One can verify that the average number of citizens is indeed 
$\sum mq_m=n$ while the average number of cities is:
\begin{equation}
Q=\frac{np|\ln p|}{1-p}.
\end{equation}
Several remarks are in order: due to the above deficit, 
the odds are always against the existing cities. If the 
deficit is small (or $p\ll 1$), fluctuations can still make 
a large city arise. The parameter $p$ sets the cutoff size 
$m^*\simeq 1/p$ of the power law behavior $q_m\sim 1/m$. 
The condition $p\ll 1$ is equivalent to the statement
that the the above deficit is small, or that the 
creation of a new city is a rare event, which seems
realistic. The distribution $q_m\sim 1/m$ is very broad. 
Inverting the rank function $R(m)$,
we find $m(R)=m^*\exp(-R+1)$, very different from the observed 
Zipf's law.

The linear model implies no interaction among citizens. 
From 
an individual's viewpoint, the chance to 
leave (or arrive at) a city is independent of the 
city's size $m$, 
everybody is free to move around. 
The simplest interaction we may consider is pair-wise 
type, 
this leads us to $w\sim m^2$. 
In this case we choose $w_a(m)=(1-p)m^2/W$ and 
$w_d(m)=m^2/W$.
The calculation is, again, standard and the result is
that
\[n=\frac{pW|\ln p|}{1-p},~~~~~~ Q=\frac{n}{|\ln p|}
\sum_{m=1}^\infty \frac{(1-p)^m}{m^2}
\simeq \frac{\pi^2 n}{6|\ln p|},\]
and 
\begin{equation}
q_m=\frac{Wp}{1-p}\frac{(1-p)^m}{m^2}.
\end{equation}
Note that $q_m\sim 1/m^2$, for $m\ll m^*\simeq 1/p$. Again using 
the rank relation,
we find $R(m)\sim 1/m$, or in the more familar form of Zipf
$m(R)\sim 1/R$. Therefore we
may draw the conclusion that it is the pairwise interaction 
that is behind the Zipf
law for city distribution, a general explanation indeed.
Recently Zanette and Manrubia have proposed a city
formation model\cite{zm}. 
They use a multiplication and diffusion process
and find that their results also reproduce the Zipf's law.
A possible motivation for such a multiplicative 
process is that citizens of the same city are subject
to the same aggregate shocks, which tend to 
increase or decrease the population size.
The implicit assumption is that the strength of
such random shocks does not depend on the size $m_i$
of city $i$.
Our present approach is complementary to theirs: 
since there are more or less 
$m^2$ departures and arrivals, the net increase (or decrease) 
$\delta m$ of the population of a city in a unit time interval, 
is {\em linearly} proportional to $m$. This leads to a 
system with multiplicative noise.
This argument shows, on one side that multiplicative noise results 
from a pairwise interaction, on the other that aggregate shocks, i.e.
random events which affect equally each citizen of a given city,
also would lead to the $m^2$ transition rates in our model.

On the other hand, the linear case $w_i\sim m$ is characterized
by fluctuations $\delta m$ which are proportional
to $\sqrt{m}$. As it has been discussed in another
reproduction-diffusion system\cite{zsp}, this is typical of
a system of non-interacting individuals.

These results are confirmed by numerical simulations. We 
show in fig. \ref{fig1} the Zipf plot of a population of cities 
obtained for $w_a(m),~w_d(m)\sim m^2$.
This compares well with actual data \cite{zipf}.

One may argue that, in reality, {\it both} linear and square 
terms should be present, since an individual's decision 
must have both 
his independent as well as interactive parts. 
Therefore it is natural to consider the mixed case. For 
the ease
of presentation we assume the transition probabilities to be:
\begin{equation}
w_a(m)=\frac{m^2}{m_0}+m\,,\;\;\;\;w_d(m)=e^{1/m^*}( 
\frac{m^2}{m_0}+a m) .
\label{mixed}
\end{equation}
The simplest way to derive the distribution of city 
sizes,
in this case, is to use detailed balance: 
The number of cities of size $m$ becoming of size 
$m-1$ is $q_{m,t}w_d(m)$. In the steady state, this has to
balance the number of cities of size $m-1$ becoming of 
size $m$, i.e. $q_{m-1,t}w_a(m-1)$.
This readily gives
\[q_m=C \frac{e^{-m/m^*}}{m}\frac{\Gamma(m+m_0)}
{\Gamma(m+1+am_0)},\]
where the constant $C$ depends on the rate $p$ at which
new cities are created.

For sizes $m\ll m_0$ the distribution is 
practically the same as that for the linear case,
$q_m\sim 1/m$. 
If the cut-off size $m^*\ll m_0$, we conclude
that the quadratic term is not relevant. If, on the other 
hand $m^*\gg m_0$, the linear behavior $q_m\sim 1/m$
holds for $m\ll m_0$ and then it crosses over to 
a power law behavior 
\[q_m\sim m^{-2-(a-1)m_0}\,\hbox{for 
$m_0\ll m\ll m^*$}.\]
The Zipf exponent is therefore $\gamma=1/[1+(a-1)m_0]$.
Zipf's law $\gamma\simeq 1$, obtains only if the pairwise
interaction is dominant, i.e. more precisely if
$|m_0(a-1)|\ll 1$. In the original Zipf's work
\cite{zipf}, the scaling law is valid only for large cities. 
The above results with $|a-1|\ll 1$, allow for a
scenario where the distribution of city sizes
crosses over from Zipf's law with $\gamma\cong 1$
for large cities $m\gg m_0$, to a non-interacting
situation $q_m\sim 1/m$ for small towns $m\ll m_0$.
The population dynamics in small towns is 
dominated by the linear term in the transition rates,
which describes non-interacting individuals. In
large cities, on the other hand, interaction
dominates and it leads to Zipf's law.
Note that Zipf's law, with a general value of $\gamma$
can also occur if the condition $|m_0(a-1)|\ll 1$ is
not met.

It is also interesting to consider a general
model with $w_a(m)=(1-p)m^\alpha/W$ and 
$w_d(m)=m^\alpha/W$. 
This indeed allows us to investigate the effects
of multi-person interactions. For example $\alpha=3$
would represent a three person interaction.
Having  found that for $\alpha=1,2$ the solution is:
\begin{equation}
q_m=\frac{pW}{1-p}\frac{(1-p)^m}{m^\alpha},
\end{equation}
we can try this solution in the equation. Treating
separately the cases $m=1$ and $m>1$ we see indeed
that this is the solution. Such a solution is also
readily found using detailed balance.
The numbers $n$ and $Q$ are given 
by
\begin{equation}
n=\frac{pW}{\Gamma(\alpha-1)}
\int_0^\infty \frac{t^{\alpha-2}dt}{e^t-1+p}.
\end{equation}
If $\alpha>2$ the integral is finite as $p\to 0$
and one has simply $n\sim pW$. For
$1<\alpha<2$ the integral diverges as $p\to 0$
and the leading term is $n\sim 
p^{\alpha -1} W$. In the same way one can 
calculate the number of cities
\begin{equation}
Q=\sum_{m=1}^\infty q_m=\frac{pW}{\Gamma(\alpha)}
\int_0^\infty \frac{t^{\alpha-1}dt}{e^t-1+p}.
\end{equation}
For $\alpha>1$ the integral is finite for $p=0$,
so that $Q\sim pW$. 

The interesting point here is that the average size of 
cities
is finite for $\alpha >2$, $n/Q\simeq \zeta(\alpha-1)/
\zeta(\alpha)$, where $\zeta(x)$ is Riemann's Zeta 
function. This observation raises the question of what
happens when the population density $\rho=n/Q$ grows
beyond the value $\rho_c=\zeta(\alpha-1)/\zeta(\alpha)$.

The answer is that the excess population concentrates
in just one city, which therefore has a finite, large
fraction of the total population. This effect has been
studied recently \cite{burda} in an equilibrium model where an
Hamiltonian $H=\sum_i \ln m_i$ was considered. The 
equilibrium distribution at inverse
temperature $\beta=\alpha$ is clearly given by a power
law distribution. However for $\alpha>2$, there are two
phases: a fluid phase for $\rho<\rho_c(\alpha)$ which
is well described by a power law distribution with an 
exponential cutoff, and a {\em condensed} phase where
a ``mega-city'' nucleates, containing a finite
fraction of the total population. It is easy to 
understand
this transition from equilibrium considerations: 
Let us consider the state in which we assign to 
each city $i$ a random number of citizens $m_i$
drawn from a power law distribution with exponent
$\alpha>2$. This state minimizes the entropy and it
has a density which is given by $\rho=n/Q=\langle{m_i}\rangle=
\rho_c(\alpha)$.
If we want to find a state with a lower density, one
can introduce a chemical potential, which is equivalent 
to
an exponential cutoff on the distribution of $m_i$.
On the other hand if one wants to build a state of 
higher density $\rho>\rho_c$ one gets into troubles. 
There are two ways out: The first is to put some 
$\delta m_i= \rho-\rho_c$ extra citizens in each city.
This results in a state which has a free energy cost
$\delta F\simeq Q\ln(\rho/\rho_c)$. The second way
out is to put all the $Q(\rho-\rho_c)$ excess citizens
in only one city. This leads to a free energy cost
$\delta F\simeq \ln[Q(\rho/\rho_c-1)]$. This only
grows logarithmically with the system size whereas the 
first variant leads to an extensive increase of the 
free energy. 
It is then clear that in the equilibrium state  the 
system prefers to create a ``mega-city'' to accommodate
the excess population. 

\begin{figure}
\centerline{\psfig{file=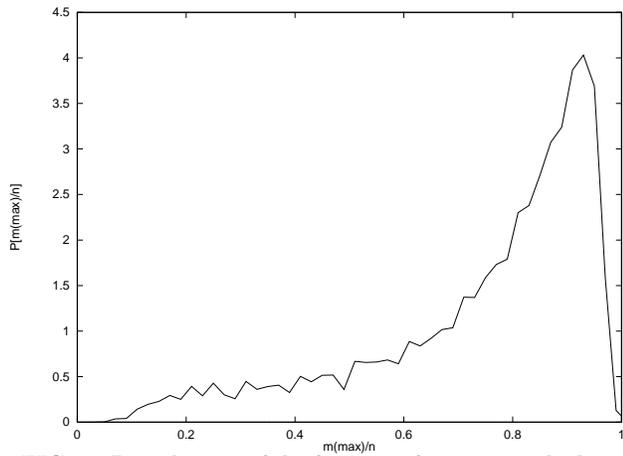,width=8.5cm,angle=270}}
\caption{Distribution of the fraction of citizens in the
largest city.}
\label{fig2}
\end{figure}

This discussion clearly refers to an equilibrium model.
Metropolis or Montecarlo dynamics of this equilibrium 
system \cite{luck} is different from
the dynamics of our master equation. Furthermore,
and more importantly, we deal with a system in which
nor the number of cities $Q$ nor the population size $n$
are fixed. In other words the density $\rho$ is not 
fixed.
For $p\ll 1$ the average density is very close to 
$\rho_c(\alpha)$. The density, no matter how it fluctuates, 
will sweep across $\rho_c$ in time. This suggests
that, in our model, dynamic nucleation of a ``mega-city''
should occur for $\alpha>2$. 
This agrees indeed with the results of numerical 
simulations. We show in fig.\ref{fig2} the distribution
of the fraction $m_{\max}/n$ of citizens which live in
the biggest city. For $\alpha=3$, the distribution is peaked at 
values  very close to $1$ (for $\alpha=2$ the critical 
density diverges $\rho_c=\infty$. Some precursor effects of 
the transition are however visible).

This leads to our second observation, that processes 
with $\alpha >2$ must result in the nucleation of
a ``mega-city'', which is not very realistic.
We can conclude that interactions of higher orders
(than pairwise) are not relevant in the
dynamics of city formation.

We see that the interaction leading to the Zipf's
law is, on one hand, the simplest possible (pairwise
interaction). On the other it is a rather special one,
since it is the ``lowest order'' of interaction which
does not lead to the formation of a mega-city, which
draws a good portion of the whole population.
The absence of a mega-city suggests that
in an expansion of the transition rates in powers of $m$,
we should neglect terms of order higher than the second.
This leads us to the mixed interaction case, which
gives very realistic results. 

In many disparate societies, it is not unnatural to assume 
that individuals make their city-dwelling 
decision based on their own opinions as well as on
its interaction with other citizens. If this indeed is the case,
we show that the larger cities obey approximately the Zipf's law.

We thank Ricard Sole for his earlier participation. This work is supported in part by
Swiss Science Foundation through the grant 20-46918.96.


\begin{references}
\bibitem{data} Data are drawn from the same sources as ref. [3].
\end{references}
\end{document}